# SpikeDeep-Classifier: A deep-learning based fully automatic offline spike sorting algorithm

Muhammad Saif-ur-Rehman[1,6], Omair Ali[1], Robin Lienkämper[1], Sussane Dyck[1], Marita Metzler[1], Yaroslav Parpaley[1], Jörg Wellmer[1], Charles Liu[2], Brian Lee[2], Spencer Kellis[3], Richard Andersen[3], Ioannis Iossifidis[4], Tobias Glasmachers[5], Christian Klaes[1]

[1]Department of Neurosurgery, University Hospital, Knappschaftskrankenhaus Bochum GmbH, Ruhr-University Bochum, Germany;
[2]Neuroresotoration Center and Department of Neurosurgery and Neurology, University of Southern California, U.S.A; [3]Division of Biology and Biomedical Engineering, CALTECH, U.S.A; [4]Institute of Informatics, University of Applied Sciences, Bottrop, Germany; [5]Institute of Neuroinformatic, Ruhr-University, Bochum, Germany; [6]Department of Electrical Engineering and Information Technology, Ruhr-University Bochum

## Abstract

***Objective.*** Recent advancements in electrode designs and micro-fabrication technology has allowed existence of micro-electrode arrays with hundreds of channels for single cell recordings. In such electrophysiological recordings, each implanted micro-electrode can record the activities of more than one neuron in its vicinity. Recording the activities of multiple neurons may also be referred as multiple unit activity (MUA). However, for any further analysis, main goal is to isolate the activity of each recorded neuron and thus called single unit activity (SUA). This process may also be referred as spike sorting or spike classification. Recent approaches to extract SUA are time consuming, mainly due to the requirement of human intervention at various stages of spike sorting pipeline. Lack of standardization is another drawback of the current available approaches. Therefore, in this study we proposed a standard spike sorter "SpikeDeep-Classifier", a fully automatic spike sorting algorithm. ***Approach.*** We proposed a novel spike sorting pipeline, based on a set of supervised and unsupervised learning algorithms. We used supervised, deep learning-based algorithms for extracting meaningful channels and removing background activities (noise/artifacts) from the extracted channels. We also showed that the process of clustering becomes straight-forward, once the noise/artifact is completely removed from the data. Therefore, in the next stage, we applied a simple clustering algorithm (K-mean) with predefined maximum number of clusters. Lastly, we used a similarity-based criterion to keep the distinct clusters and merge the similar looking clusters. ***Main results.*** We evaluated our algorithm on a dataset collected from two different species (humans and non-human primates (NHPs)) without any retraining. Data is recorded from five different brain areas with two different recording hardware and three different electrodes types. We showed that the "SpikeDeep-Classifier" has the potential of being automated and reproducible spike sorting algorithm, universally. We compared the result of algorithm with ground-truth labels. We also validated our algorithm on publicly available labeled dataset. ***Significance.*** The results demonstrated that the SpikeDeep-Classifer provides a universal solution to fully automatic offline spike sorting problem.

# Clinical trial registration number

# Introduction

Complex behaviors and understanding network properties in the brain requires access to the activities of population of neurons. This necessary access is mostly provided by implanting small but dense micro-electrode arrays. State-of-the art single micro-electrode array contains hundreds of channels which enable us to record SUA of hundreds of neurons (Lambacher, et al., 2011; Spira & Hai, 2013; Berényi, et al., 2013; Frey, Egert, Heer, Hafizovic, & Hierlemann, 2008; Harris, Quiroga, Freeman, & Smith, 2017). Now a days, it is a common practice to implant multiple micro-electrode arrays and record neural activities from more than one cite, simultaneously (Aflalo, et al., 2015; Klaes, et al., 2015). However, recorded data is usually contaminated with on-neural activities (noise/artifacts), and in addition to that it is also possible that a single channel records the activities of more than one neuron (MUA). Thus, the biggest constraint for further analysis is to extract and isolate the activity of each single neuron (SUA) in the presence of background activities and MUA. This process is called spike sorting.

The process of spike sorting is usually either manual or semi-automatic (Abeles & Goldstein, 1977; Gibson, Judy, & Marković, 2012; Lewicki, 1998). Process of manual or semi-automatic spike sorting involves human involvement at various stages of spike sorting pipeline. As a result, this process become labor-intensive and highly time-consuming. Therefore, manual or semi-automatic spike sorting technique could never compete with increasing amount of data resulting from highly dense micro-electrode arrays and long duration recording sessions (Berényi, et al., 2013). Another major drawback of involving humans in this loop is that they are inconsistent about their decisions. There could be considerable variability in labeling of data recorded on different days across different sessions (Wood, Black, Vargas-Irwin, Fellows, & Donoghue, 2004). Another limitation of manual or semi-automatic sorting is that the quality of spike sorting is completely dependent on the skills of human curator. Therefore, fully automatic spike sorting has always been major area of interest (Spacek, Blanche, & Swindale, 2009; Takekawa, Isomura, & Fukai, 2012; Bongard, Micol, & Fernández, 2014; Carlson, et al., 2014; Pachitariu, Steinmetz, Kadir, Carandini, & Harris, 2016; Yger, et al., 2018; Chung, et al., 2017).

Mostly, a spike sorting pipeline involves at first the pre-processing of the raw time series by applying band-pass filtering and then using threshold to extract qualified events. It is possible that few of the extracted qualified events represent background activities and others represent either SUA or MUA. Finally, to assign labels to each of the extracted qualified events clustering is used (Lewicki, 1998; Einevoll GT, 2012; Marre, et al., 2012). Usually, at least one of these processes is performed manually.

It has been reported before that, a considerable fraction of channels of implanted array records only background activities (Lewicki, 1998; Hill, Mehta, & Kleinfeld, 2011; Klaes, et al., 2015; Rey, Pedreira, & Quiroga Quian, 2015). Contrarily, other channels which record SUA also record substantial amount of background activities. These background activities are generated with the combination of technical artifacts and neural activities that are situated far away from the tip of recording electrodes. Position of recording electrodes can also be slightly perturbed by slight movements. Therefore, resulting signal is of non-stationary nature (time variant). The dynamics of recorded signal changes from session to session. Therefore, it is a challenge to model all the resultant dynamics.

Recently, it has been reported (Chung, et al., 2017) that unavailability of general solution to spike sorting is mainly because of non-stationary behavior of background activities. In this study, we showed that the spike sorting becomes an ordinary clustering problem upon the complete removal of background activities from the source signal. We also proposed a universal solution for this problem, based on our previous study (Saif-ur-Rehman, et al., 2019) in conjunction with a novel algorithm called 'Artifact rejector'. By universal, we mean that the algorithm is trained once on a versatile dataset. Later, it can be evaluated on a dataset recorded from a different brain area, different species, different electrode type and recording hardware without any re-training. We showed that it is possible to completely remove background noise with huge amount of labeled training data and by stacking two different deep-learning methods.

Deep-learning methods, specially supervised learning algorithms in combination with huge number of labeled training examples have proven worthwhile in the field of computer vision (Krizhevsky, Sutskever, & Hinton, 2012). Recently, Convolutional neural networks (CNNs) alone have become the major source of success in many computer vision applications (Guo, Dong, Li, & Gao, 2017). CNNs, because of shared-weights architecture and translation in-variance characteristics can learn temporal and spatial patterns (LeCun, Bottou, Bengio, & Haffner, 1998). However, the primary reason behind the success story of CNNs is the availability of huge publicly available datasets (Jia, et al., 2009; Stallkamp, Schlipsing, Salmen, & Igel, 2011). We strongly believe that results of many neuroscience problems including spike sorting and online brain computer interface (BCI) decoding can be improved in the presence of huge labeled datasets. Therefore, in this study, we collected and labeled a huge dataset. Dataset includes the data from our own lab and from different collaborators. We also used some publicly available labeled datasets to validate our results (Shi, Apker, & Buneo, 2013; Lawlor, Perich, Miller, & Kording, 2018; Perich, Lawlor, Kording, & Miller, 2018; Buneo, Shi, Apker, & VanGilder, 2016).

In this study, we aimed to provide a universal solution to offline spike sorting problem by making full use of huge labeled dataset and deep learning algorithms. Our algorithm 'SpikeDeep-Classifier' is based on a novel pipeline, which is a set of supervised learning methods and unsupervised learning methods. First supervised learning method is used to select the meaningful channels as proposed in (Saif-ur-Rehman, et al., 2019) . Then, we employed another supervised learning method to remove remaining non-neural activities from the selected channels. After the complete removal of non-neural activities, we employed a K-Mean clustering (Lloyd, 1957; Macqueen, 1967) with predefined number of maximum clusters on the feature vectors extracted using Principal component analysis (PCA) (Jolliffe & Cadima, 2016). Finally,, a similarity measured based algorithm is used to automatically accept distant clusters and merge similar looking clusters.

## Materials & Method
### Approvals
We used a dataset collected from two tetraplegic patients implanted with two Utah arrays each and the epilepsy patients implanted with depth-electrodes in preparation for surgery. Utah arrays patients were implanted in posterior parietal cortex (PPC). These Patients were recruited for two different BCI studies (Aflalo, et al., 2015; Klaes, et al., 2015). These studies were taken place after the institutional approvals held by California Institute of Technology, and University of Southern California. Detailed approvals information is available in (Aflalo, et al., 2015; Klaes, et al., 2015). Epilepsy patients were implanted with depth-electrodes/ micro-wires in hippocampus in the form of bundles.  These patients were implanted for medical reasons and have participated voluntarily. We obtained the approval for epilepsy patients from Ruhr-University ethics committee. In addition, we also used publicly available datasets. Approval of each dataset is available in (Lawlor, Perich, Miller, & Kording, 2018; Shi, Apker, & Buneo, 2013).

## Demographic and Implantation Details

In this study we used the data recorded from five human patients and four NHPs (male rhesus macaques).

Human patients were implanted either with Utah arrays or with micro-wires using a Behnke-Fried configuration (Fried, et al., 1999). Three of human patients were implanted with micro-wires, which were coupled in a group of 8 individual platinum coated electrodes. Remaining two human patients were implanted with two Utah arrays. Each array contains 100 electrodes arranged in a grid of dimension 10x10.Further information about surgery and array placement is mentioned in (Aflalo, et al., 2015; Klaes, et al., 2015).

For NHPs, we used two different publicly available datasets provided by Collaborative Research in Computational Neuroscience (CRCNS) (Perich, Lawlor, Kording, & Miller, 2018; Buneo, Shi, Apker, & VanGilder, 2016). Dataset reported in (Lawlor, Perich, Miller, & Kording, 2018) recorded from two macaques implanted Utah arrays. Another dataset (Shi, Apker, & Buneo, 2013) is recorded from two rhesus macaques (macca mulatta) using single micro-electrodes.

Further demographic and implantation details are mentioned in Table 1.

*Table 1: Subjects Demographic and implantation details*

| Specie | Subject ID | Sex | Age (Year) | Place of Implantation | Number of Recordings | Number of Implanted electrodes |
|---|---|---|---|---|---|---|
| Humans | U1 | Male | 32 | Posterior parietal cortex | 10 | 192 (2-Utah array) |
| Humans | U2 | Male | 63 | Posterior parietal cortex | 10 | 192 (2-Utah array) |
| Humans | M1 | Female | 49 | Anterior hippocampus | 1 | 16 (Micro-wires) |
| Humans | M2 | Female | 26 | Anterior hippocampus | 1 | 16 (Micro-wires) |
| NHPs | MM | Male | - | Primary motor cortex and premotor cortex | 1 | 192 (2-Utah array) |
| NHPs | MT | Male | - | Primary motor cortex and premotor cortex | 1 | 192 (2-Utah array) |
| NHPs | X/ B | Male | - | Superficial cortex | 10 | Single micro-electrodes |

## Data Collection & Preprocessing

Humans data is recorded using a Blackrock microsystems neural signal processor (NSP). Here, we aim for end-to-end learning (Glasmachers, 2017; LeCun, Bengio, & Hinton, 2015) first for meaningful (neural) channel selection and then to discard background activities. Therefore, we minimally preprocessed the raw data. Preprocessing involves extraction of events (spike candidates) from the raw data based on thresholding procedure (Lewicki, 1998). The threshold setting of event detection was set to -4.5 times the root-mean-square (rms) of high-pass filtered signal with cutoff frequency 250 Hz. We used similar setting in our previous study (Saif-ur-Rehman, et al., 2019) for meaningful channel selection and also used in (Klaes, et al., 2015) for online BCI decoding. Each extracted event is a 48 sampled waveform, 15 sample before threshold and 32 after threshold. Total duration of an extracted event is 1.6 ms.

We also used two different publicly available datasets (Perich, Lawlor, Kording, & Miller, 2018; Buneo, Shi, Apker, & VanGilder, 2016) to validate findings of this this study. Dataset (Perich, Lawlor, Kording, & Miller, 2018) contains data from two NHPs. Data was recorded using Blackrock microsystems NSP. Dataset contain 48 samples preprocessed labeled events. Further details are available in (Perich, Lawlor, Kording, & Miller, 2018). Second dataset (Buneo, Shi, Apker, & VanGilder, 2016) is recorded with Plexon Inc NSP. Dataset contain 32 sampled events, which were than resampled to 48. Further information is available in (Buneo, Shi, Apker, & VanGilder, 2016).

### Data labeling

Proposed spike sorting pipeline is a set of supervised and unsupervised learning algorithms. Supervised learning is gradient-based and tries to minimize the defined cost function by comparing predicted output and true output. Therefore, labeled training data is required. We labeled the given event either as 'Spike' or 'Artifact'. Events representing action potentials (neural activities) are labeled as spike. Contrarily, events representing background activities (muscle artifacts, noise) are labeled as artifacts. Process of labelling is done in a semi-automatic way using Gaussian mixture model (GMM) and careful visual inspection. Further detail about labeling process is being explained in our recent study (Saif-ur-Rehman, et al., 2019).

### Data distribution for training and validation

We used the pretrained model of SpikeDeeptector for meaningful channels selection. Other supervised learning method in the proposed spike sorting pipeline is Non-neural rejector (NNR). NNR takes a single event from the meaningful channel and predict it as 'Spike' or 'Artifact'. For training the NNR, we consider data from human patients and from NHPs. We considered four recording sessions from each patient implanted with Utah arrays and one recording session from a patient implanted with microwires. From NHPs, we considered a data from one recording of a subject (MT) implanted with Utah array. We also used data of 6 random days of subjects implanted with single micro-electrode. Distribution of data from all the sources is shown in **Figure 1**(a). **Figure 1** (b) shows the distribution of data from each class. There is way more training examples of class 'Spike' as compared to class 'Artifacts'. Therefore, to avoid any bias during training, we considered 400,000 training examples from each class.

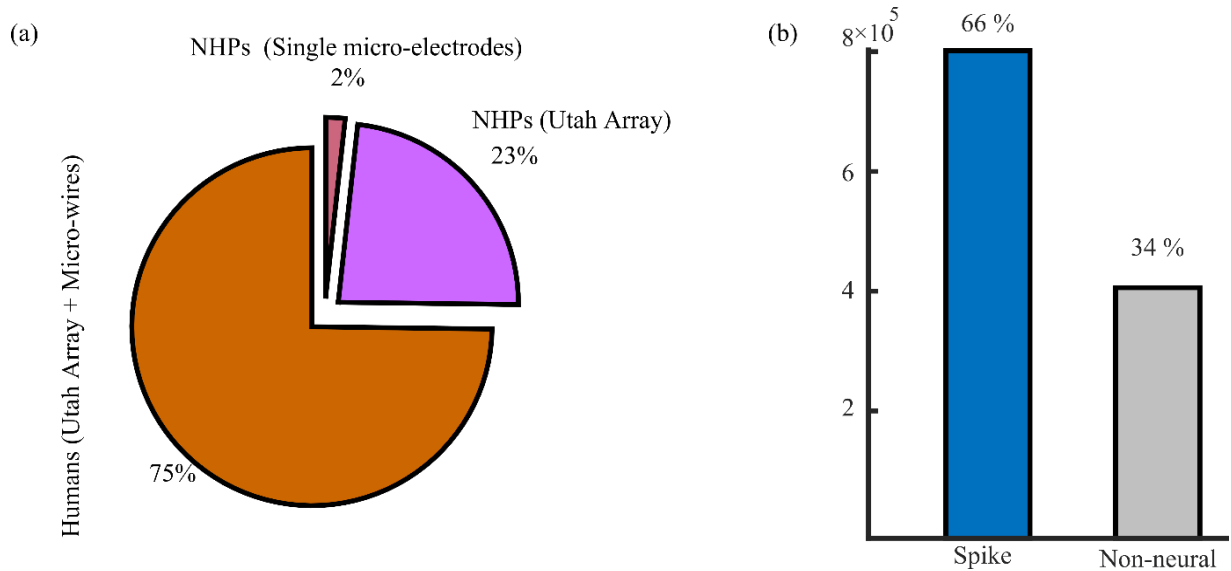

*Figure 1: Training data distribution. (a) Contribution of each source in training dataset. (b) Number of feature vector of each class.*

## SpikeDeep-Classifier algorithm

In this study, we proposed an offline automatic spike sorter called SpikeDeep-Classifier. SpikeDeep-Classifier is a combination of supervised and unsupervised learning algorithms as shown in **Figure 2**. We completely removed background activities by stacking SpikeDeeptector (Saif-ur-Rehman, et al., 2019) and NNR, both algorithms are based on supervised learning principles. Therefore, they required labeled training data to optimize learnable parameters, iteratively. We showed that the background activities or non-neural activities can be completely removed by exposing supervised learning algorithms to a versatile dataset during training. As a result, background activities can be completely removed and more importantly this fact is valid universally. By universally, we mean by concatenating the pretrained models of SpikeDeeptector and NNR background activity can be removed from the data recorded from different brain areas, of different species, implanted with different types of sensors (micro-electrodes), using different hardware. Extraction of meaningful data is critical in neuroscience and its applications, including BCI applications and spike sorting. In study (Noc, Ciancio, & Zollo, 2018), it has been shown that spike detection is the first and most pivotal step in neuro-prosthetic applications. Next stage in this pipeline is dimensionality reduction of events correspond to neural data using PCA, which is one of the standard algorithms for dimensionality reduction in spike sorting applications. Here, instead of using first two principal components for clustering, we defined a criterion that keeps most of variability in data and give us resulting principal components. We then employed a clustering algorithm on the extracted feature of events representing neural data. We showed upon the removal of background activity using NNR, spike sorting can be done with a very simple clustering algorithm e.g. K-Mean with predefined maximum number of clusters. Later, we defined a similarity-based criterion to merge the similar looking clusters and accept the distinct clusters as a separate unit. We evaluated SpikeDeep-Classifier on different recording sessions of different subject using trained model of supervised learning algorithms and same, fixed values of other two parameters (Expected maximum number of clusters & Similarity threshold) of pipeline. The results show that SpikeDeep-Classifie provides the accuracy comparable to a human expert.

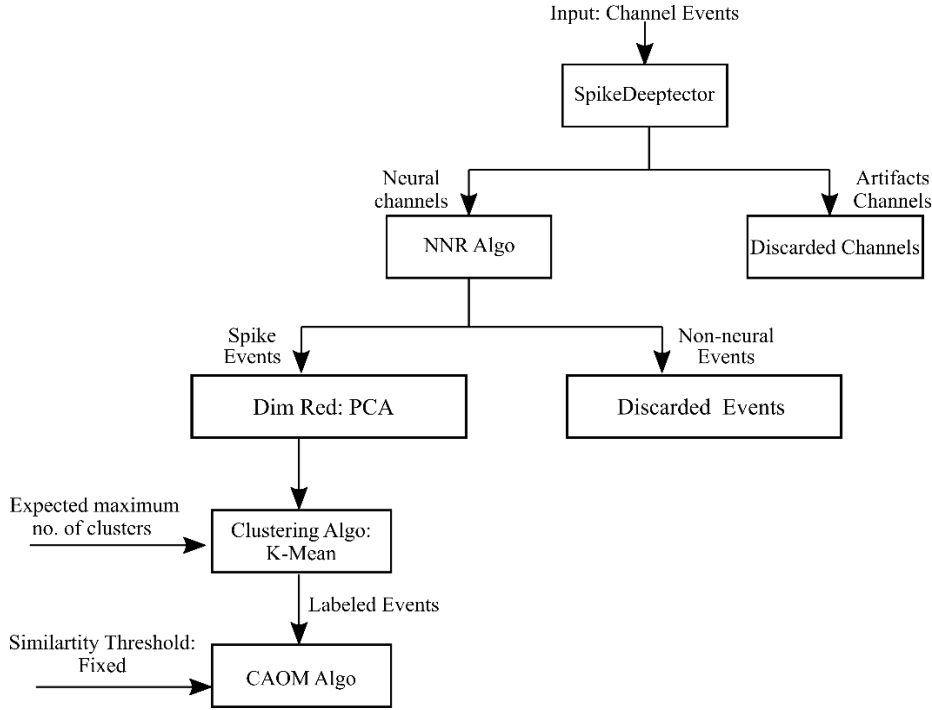

*Figure 2: SpikeDeep-Classifier Pipeline. In the first stage SpikeDeeptector selects meaningful channels. These channels are then further processed by Non-neural Rejector (NNR) to extract neural events. Extracted neural events are then projected in a low-dimensional space using PCA. We used a criterion that keeps intact 85% variability of data and reduced the dimension of inputs (neural events). Clustering is then applied on the extracted feature space with predefined number of clusters. Later, cluster accept or merged (CAOM) algorithm, automatically accept distinct clusters and merged similar looking clusters.*

## SpikeDeeptector

The process of mapping raw signals into decision space is shown in **Figure 2**. SpikeDeeptector is the first building-block of the SpikeDeep-Classifier pipeline. Goal of the SpikeDeeptector is to select the channels recording neural data and discard the channels recording only noise/artifacts. We have shown in (Saif-ur-Rehman, et al., 2019) that SpikeDeeptector can do such discrimination, universally. Therefore, in this study we used a pretrained model of the SpikeDeeptector. The process of extracting feature vectors, architecture, and process of training SpikeDeeptector is explained in detail in our previous study (Saif-ur-Rehman, et al., 2019). We extracted the feature vectors with batch size equals 20. For all the experiments, we used pre-trained model of SpikeDeeptector. The trained model of SpikeDeeptector is available on github and can be downloaded using the following link. https://github.com/saifhanjra/SpikeDeeptector/tree/master/EvaluateTrainedModel

## Non-neural Rejector (NNR)

SpikeDeeptector provides the list of channels recording neural activities. These channels in addition to neural activity also record considerable amount of background activities (non-neural activities). Therefore, we aimed to detect and discard all the event corresponding background activities from the channels recording neural activities (see **Figure 4**). Non-neural rejector (NNR) can be used as preprocessing step for a clustering algorithm. It can isolate the overlapping event corresponding to spike data and non-neural data (see **Figure 4** (b)). As a result, it provides an ease to further steps for spike sorting as shown in **Figure 4**( c). In this case, spike sorting becomes an ordinary clustering problem after the removal of overlapped events.

To achieve this goal, we made use of available labeled training data and designed a supervised learning based on standard architecture of convolutional neural networks (Krizhevsky, Sutskever, & Hinton, 2012; Guo, Dong, Li, & Gao, 2017) as shown in **Figure 3**. CNNs use shared weight which enable translation invariance and as a result produced more generic feature. These learned features are also robust against time delays and advancements in spike occurrence. Here, we used 1-D CNNs because we are interested to learn temporal pattern, only.

NNR take an input from 48 sampled event, and process it through 3 convolutional layer, two pooling layers, a fully connected and finally classifies it as a 'Background activity' or as a 'Spike' using a Softmax classifier, as shown Figure 4. At each convolutional layer, each kernel is convolved across the width of input volume and then slides with stride=1. This results in 1-D in convolved feature maps. Then, non-linearity is introduced using an activation layer. Here, we used Rectified linear units (ReLUs) $f(x) = \max(x, 0)$ (Nair & Hinton, 2010) . Except for the first convolutional layer, each conv layer is followed by polling layer. The goal of pooling is to discard the unnecessary information. Here, we used max polling. The size of each kernel and pooling layer is mentioned in **Figure 3**(b).

We tried to minimize regularized cross-entropy cost and added L2 regularization term in cross-entropy cost function. In addition, we also used batch normalization to standardize intermediate outputs of background-activity rejectors to zero mean and unit variance for the training inputs equal to mini-batch size. We used the same optimization algorithm (mini-batch gradient decent with momentum) and tuned the hyperparameters in a same way as reported in our previous study (Saif-ur-Rehman, et al., 2019).

We trained a robust NNR using the data of 2 species, 5 subjects, 6 brain areas, 3 different types of electrodes and two recording hardware. The distribution of training data is shown in **Figure 1** (b).

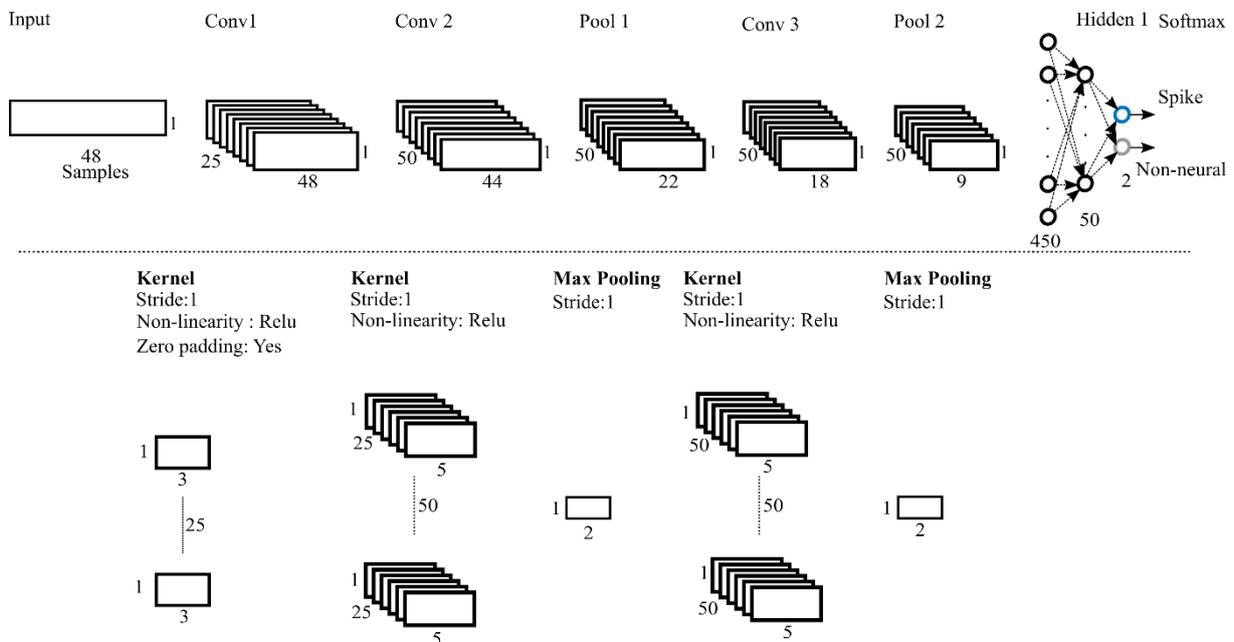

*Figure 3: Architecture of NNR. Process of mapping an input space to corresponding decision space. The input is convolved and down sampled using convolutional and pooling layers. Size of and number of kernels of each convolutional layer is also mentioned in figure. We used zero padding during the first convolutional layer only. Kernel size of pooling is also mentioned in Figure.*

## Dimensionality Reduction

Dimensionality reduction is usually performed using unsupervised learning algorithms. Mostly, spike sorting algorithms tries to eliminate redundant features and construct feature vectors for clustering algorithms. PCA is the one of the most practiced algorithm by the community (Adamos, Kosmidis, & Theophilidis, 2008; Lewicki, 1998; Souza, Lopes-dos-Santos, Bacelo, & Tort, 2018). PCA construct low dimensional feature vectors by doing eigen value or singular value decomposition of covariance matrix constructed from the presented data. Mostly, spike sorting project the high dimensional events in to corresponding 2 or 3-dimensional principal component space using the eigen vectors. PCA algorithm ensures that this 2 or 3-dimensional capture highest variability in the presented data. However, it is possible this low dimensional projection does not carry necessary discrimination power. Therefore, in this study we presented a criterion that keep intact certain amount of variability of the presented data and construct low dimensional feature vectors. Here, we used a criterion that keeps intact 85% of variability of data.

## Clustering Method

One of the cornerstones of this study is to show spike sorting becomes an ordinary clustering problem upon the removal of background activity. Usually, neural data generated further away from the tips of recording electrodes and background activities are overlapped as shown in Figure 4 (a, b). Therefore, it is very hard for any clustering algorithm to predict them as separate clusters. However, supervised learning algorithms are very powerful models. Particularly, deep learning algorithms can even learn hidden patterns. For this reason, we trained a deep learning algorithm to isolate background activities, as shown Figure 4 (b). After the removal of background activities clustering become trivial as shown in Figure 4 (c). Two clusters are quite distinct from each other and are quite distinguishable. Even a simplest clustering algorithm like K-mean can perform well as shown in Figure 4.

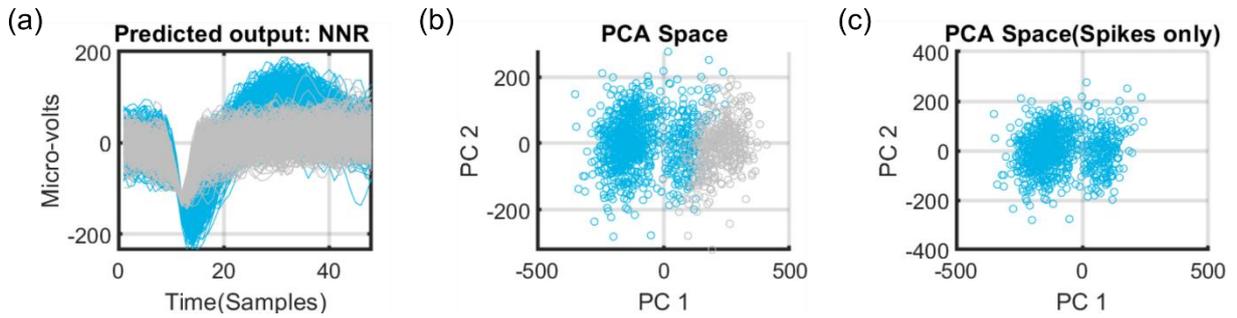

*Figure 4: Non-neural rejector goal. (a) predicted output of NNR. Events (Blue) are predicted as spike and events (grey) are predicted as non-neural. (b) projection of corresponding events in 2-dimensional space using PCA. Few events corresponding to class spike and non-neural are overlapped. (c) shows NNR isolates the overlapping events and provide an ease for later processing (clustering).*

We used K-Means as a clustering algorithm with squared Euclidean distance metric and k-means++ algorithm for defined number of clusters centers initialization. K-mean is a type of unsupervised learning algorithm of clustering unlabeled data into k-clusters. We defined k as maximum number of expected clusters, equals to either 3 or 4. Then, later accept the clusters or reject the clusters based on the defined criteria defined in a section (Automatically Accept or Reject clusters).

## Cluster Accept or Merge Algorithm (CAOM)

We used K-mean clustering and defined K as a maximum number of expected clusters in one channel. During data labeling, we observed maximum number of clusters in one channel is three, which is also rare. Here, we introduced a very simple method to accept the distinct cluster and merge the similar looking clusters. We are considering the data of multiple species, recorded from different brain areas using different recording hardware and different types implanted electrodes. It is possible that recorded data can be on different scales. Therefore, we first normalized the data using z-normalization (Patro & Sahu, 2015) as a preprocessing step. Z-normalization ensures, that all the features have zero mean and standard deviation equals to one. Then, we measured similarity between each cluster. Hence, we compared mean Euclidean distance of each cluster. We either merged two clusters with minimal distance less than the defined threshold distance, or all the defined clusters are accepted and are originated from independent sources. In the case of merging two clusters, new mean of merged clusters is calculated and compared with remaining clusters mean. This process of merging clusters is repeated unless mean Euclidean distance of each cluster from each other is greater than defined threshold.

Threshold distance is the only hyperparameters of CAOM algorithm. We tuned value of 'threshold distance' by visual inspection. We fix the value of threshold distance to 5.5 after tuning it. Later, we used the same value for all recording sessions used for validation.

# Results

SpikeDeep-Classifier is a pipeline that presents a solution to spike sorting problem. We evaluated the SpikeDeep-Classifier pipeline on the data of three human patients and two NHPs. Two human patients were implanted with Utah arrays and for each patient three recording sessions were considered; third human patient was implanted with microwires and only one recording session was available. Similarly, two NHPs were either implanted with two Utah arrays or single microelectrode. One Recording session of Utah arrays subject and four recording sessions another subject is considered for evaluation.

## SpikeDeeptector

SpikeDeeptector is the first component of SpikeDeep-Classifier pipeline. Main aim of SpikeDeeptector is to isolate the channels containing neural activities. So, that for any further processing only the channels containing neural activities are considered and the remaining channels are discarded.

We evaluated previously trained model of SpikeDeeptector on the data of four subject groups and multiple recording sessions, as shown in **Table 2**. SpikeDeeptector has wrongly classified only 3 channels out of 692 channels, which shows SpikeDeeptector has a good quality of generalization. We also highlighted the consistent performance of SpikeDeeptector by evaluating it on each session, individually. This evaluation aspect is evident on all three different types recording sessions with few, some and several channels recording neural activities. Evaluation performance of SpikeDeeptector on each type of recording session is reported in (**Supplementary Material: SpikeDeeptector**, **Table 6**).

*Table 2: Cumulative Performance evaluation of SpikeDeeptector on the data recorded from human patients and NHPs during multiple recording sessions.*

| Subject Group | Number Of sessions | Neural channels (Neural Channels /Total Channels) | Artifact Channels (Artifacts Channels /Total Channels) | False Positives | False Negatives |
|---|---|---|---|---|---|
| Utah Array (Humans) | 6 | 109/576 | 467/576 | 2 | 0 |
| Microwires (Humans) | 1 | 15/16 | 1/16 | 0 | 0 |
| Utah Array (NHP) | 1 | 95/96 | 1/96 | 0 | 1 |
| Single Microelectrodes (NHP) | 4 | 4/4 | 0 | 0 | 0 |

## Non-neural Rejector (NNR):

SpikeDeeptector provide the channels recording neural data. However, channels recording neural events can also record the non-neural events. Therefore, in next stage we aimed to isolate these two kinds of events and discard the events correspond to non-neural activities. We trained NNR rejector on the training data. Further details of training dataset are explained in section (**Data distribution for training and validation**). NNR is a supervised learning algorithm. Therefore, we trained it on a labeled dataset and tried to minimize cross-entropy cost. Further, training and validation details can be seen in section (**Non-neural Rejector (NNR)**)

**Human Subjects:** We evaluated a trained model of NNR on the data of seven recording session of three human Patients. Subjects were implanted with either Utah arrays or microwires on different areas of brain (PPC or Hippocampus). We reported evaluation accuracy in a confusion matrix as shown in **Table 3**. NNR provides 93.4% accuracy on the feature vectors of class 'Spike' and 86.4 % accuracy on the feature vector of class 'Non-neural'. Data distribution between two classes is unbalanced with 83.6% of data represent class 'Spike' and 16.4% of data represent class 'Non-neural'. Overall classification accuracy is 92.3%. In addition to cumulative performance on the data of all human subjects, we also reported the evaluation performance of NNR on each recording session, individually (see section **Supplementary Material: Non-neural Rejector**). Performance of NNR during each individual recording session remains consistent, as shown in **Table 7** and **Table 8** with minimum and maximum reported accuracy of 88.9% and 95.4%.

*Table 3: Evaluation performance NNR on the data recorded from human patients implanted with Utah arrays and microwires. Confusion matrix reports overall accuracy and the classification accuracy of each class.*

**Non-neural Rejector**

|  | Spike (True) | Non-neural (True) |  |
|---|---|---|---|
| **Spike (Predicted)** | 384370 / 78.1% | 11029 / 2.2% | 97.2% |
| **Non-neural (Predicted)** | 27041 / 5.5%% | 70096 / 14.2% | 72.2% |
|  | 93.4% | 86.4% | **92.3%** |

Predicted Labels / True Labels

**Visualization:** We selected three different examples for visualization (see **Figure 5**). **Figure 5** shows the waveforms with associated ground truth labels and predicted labels along with mean waveforms of each class and projection of waveforms in 2-D using PCA. For visualization, we showed the response of NNR on three different kind of recording channels. **Figure 5**(a) shows the response of NNR on the channels where spike events and non-neural events are only partially overlapped. Therefore, better performance from a clustering algorithm is expected. However, in **Figure 5** (b) spike events and non-neural events are almost overlapped. Hence, it is a difficult task for a clustering algorithm to discriminate two clusters. **Figure 5** (c) show another type of channel which records only few events correspond to non-neural activity. These few events hardly present a cluster. In all the above explained conditions NNR performs equally good (or even better) in comparison with ground truth labels.

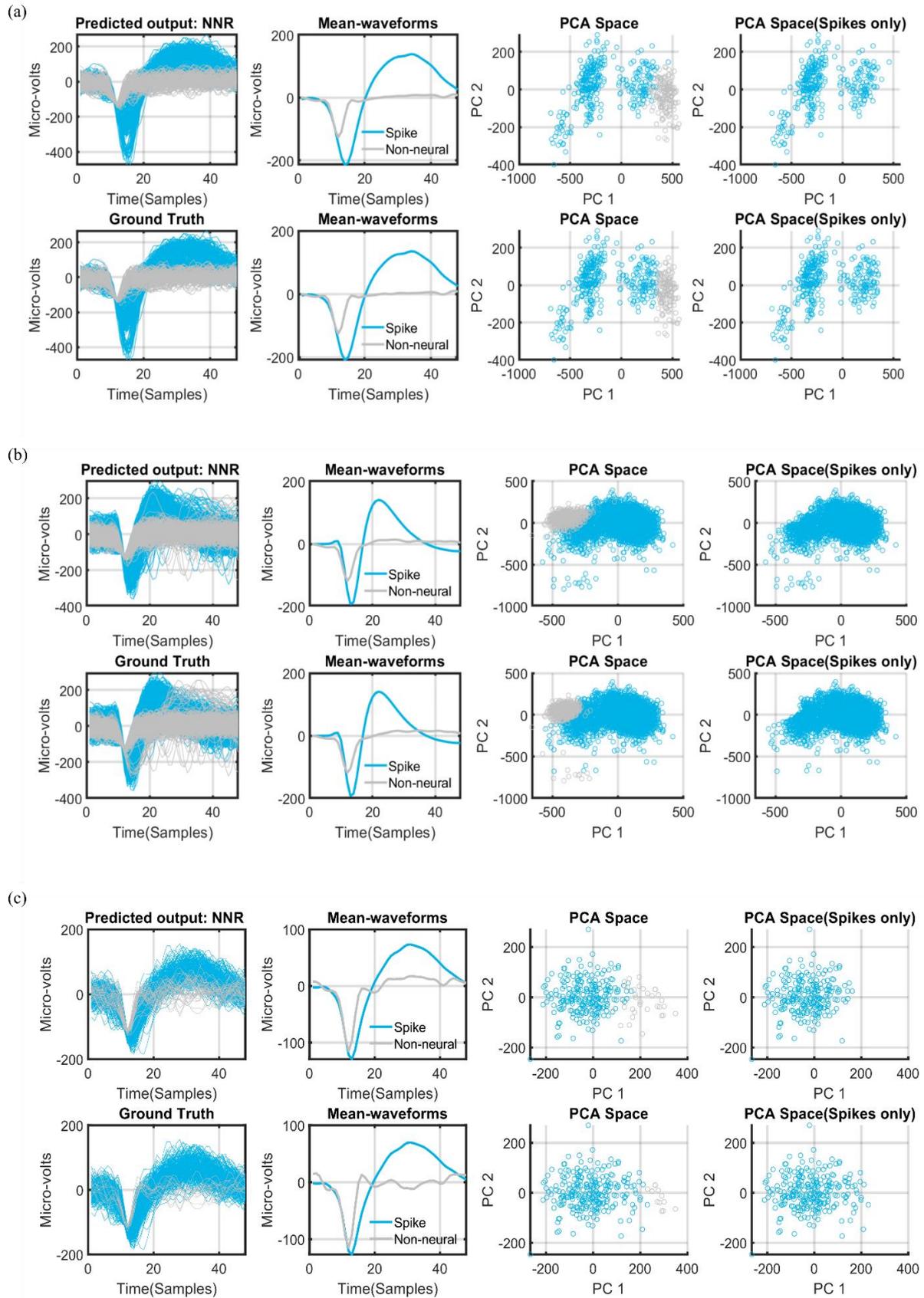

*Figure 5: Visualization examples of NNR in three different cases. (a) non-neural activities and spike activities are partially overlapped. (b) non-neural events and spike activities are completely overlapped. (c) Partially overlapped but only few events represent class artifacts. In all cases, first row shows the events with predicted labels, mean waveform of each predicted class, projection in 2-D using PCA, and 2-D projection of events predicted as spike. Second row shows the events with ground truth labels, mean waveforms and corresponding 2-D projections.*

## Significance of NNR: Overlapping Waveforms

Events represent spikes and non-neural activities can completely overlap shown in **Figure 6**(a). As a result, even humans can make mistakes during labeling (see **Figure 6** (a)), here during labeling human missed a distinct unit by merging it with non-neural activity. However, by considering NNR as a preprocessing step before clustering. The process of clustering become trivial as shown in **Figure 6** (b). Inclusion of NNR in SpikeDeep-classifier pipeline isolates the overlapped clusters. Thus, spike sorting can become an ordinary clustering problem (see **Figure 6**).

We witnessed quite few channels in one recording session with events representing spike and non-neural activities are either completely or partially overlapped. However, by employing NNR as a preprocessing step of clustering. The process of clustering become trivial. We presented few more examples in (**Significance: Non-neural Rejector: Figure 11, Figure 12**). Presented examples show that NNR can facilitate clustering process by removing overlapped events.

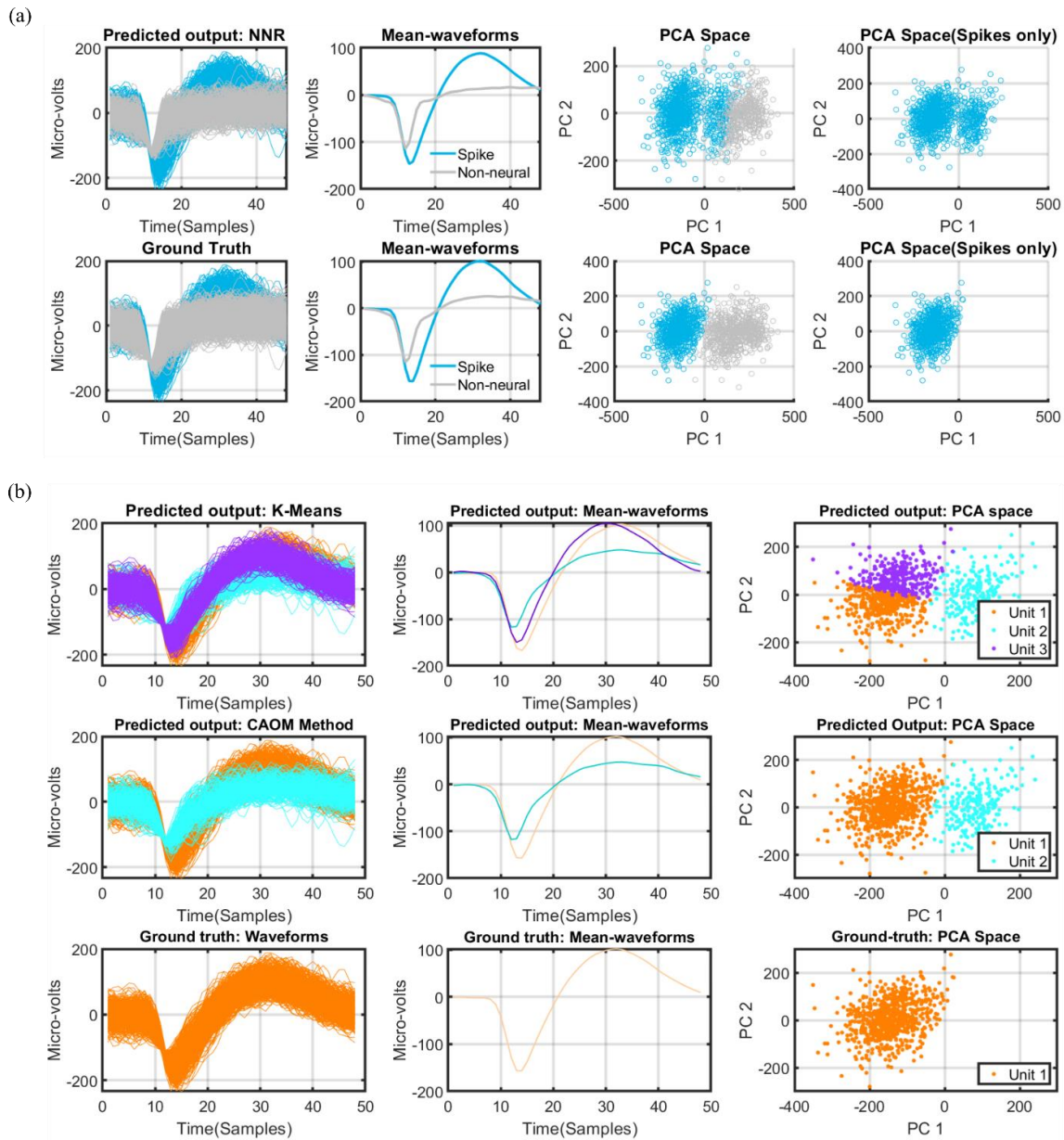

*Figure 6: Significance of NNR (Visualization). (a) Response of NNR along with ground truth labels (human). NNR has clearly outperformed human in this case. Here, non-neural cluster and spike cluster was completely overlapped. Therefore, even human misclassify some of the events and can missed a neural unit. (b) Shows the result of clustering and CAOM. SpikeDeep-Classifier pipeline outperforms human spike sorter because NNR isolates overlapped clusters. Mean waveforms of both clusters (see Result CAOM) clearly show two distinct neural units. However, human curator missed one cluster (see ground-truth).*

## Clustering & CAOM:

SpikeDeeptector in conjunction NNR completely remove non-neural activities in two steps. After the complete removal of non-neural activities, remaining data (spikes) is used to identify number of neural units present on a single channel. This process is taken place in two steps, the first step involves the process of clustering with predefined maximum number of clusters and in the second step similarity between each cluster is measured as explained in section (**Cluster Accept or Merge Algorithm (CAOM)**). Similar looking clusters are merged, and distinct looking clusters are treated as separate clusters (units). We have defined a maximum number of clusters as 3 and the similarity threshold as 5.5. We have fixed these two parameters and used it for all evaluation data.

**Humans: Utah Array Subjects:** We used K-mean clustering (see section **Clustering Method**) and then CAOM (see section **Cluster Accept or Merge Algorithm (CAOM)** to accept or merge the clusters. We evaluated our methods on six recording sessions of human patients implanted with Utah array. Out of 576 channels only 109 channels were predicted as neural channels. Two predictions were false positives. Majority of channels either record one neural source or two neural sources on a channel (see **Table 4**). However, there were few channels, where three neural sources were recorded. (See **Table 4**). K-mean clustering in conjunction CAOM has been able to predict right number clusters on most of the channels. Out of 107 channels only 8 channels were predicted with different number of clusters as labeled. Rand index is used to report the quality of clustering method. Rand index is a measure of similarity between two data clustering methods. The value of rand index is between 0 and 1. The value '1' means both clustering (ground truth, predicted) methods produced exactly same results, and 0 indicate that two data clustering methods completely disagree with each other. The achieved mean rand index is more than 0.8 for any number of neural units on a channel. Similarly, achieved mean accuracy for any number unit is more than 87%.

*Table 4: Performance evaluation of clustering method & CAOM on the data of six recording sessions of human patients implanted with Utah arrays.*

| Number of Units | No. of Channels (True) | No. of Channels (Pred) (Correct, wrong) | Rand Index | Accuracy (%) |
|---|---|---|---|---|
| 3 | 18 | (18,1) | 0.84 ± 0.08 | 87.76 ± 5.17 |
| 2 | 47 | (44,5) | 0.85 ± 0.08 | 89.09 ± 6.69 |
| 1 | 42 | (37,2) | 0.81 ± 0.12 | 87.25 ± 11.01 |

**Table 4** shows the cumulative performance of Clustering & CAOM on all the recording session of human patients implanted with Utah array. In addition to that we also showed the performance of Clustering & CAOM on all individual recording sessions (see **Supplementary Material: Clustering & CAOM**: **Table 9**). These recording sessions have different number of channels with different number of units. Performance of Clustering & CAOM remains consistent during all individual recording sessions (see **Table 9**).

**Visualization:** we also presented an example for visual inspection with three neural unit on a channel (see **Figure 7**). **Figure 7** (a) shows the predicted output of k-mean clustering algorithm with 3 clusters. Output of CAOM is shown in the second stage (**Figure 7**(b)). Here similarity between each unit is calculated based on the criteria explained in (**Cluster Accept or Merge Algorithm (CAOM)**). All three clusters have been considered as distinct clusters. Figure 6 (c) shows the ground truth waveforms, ground truth means that the waveforms and the ground projection in 2-D. **Figure 7** (b) and **Figure 7** (c) look quite similar, which shows high quality of SpikeDeep-Classfier pipeline.

In addition to example with three units on a channel, as shown in **Figure 7**. We also presented an example with two clusters (**Supplementary Material**: **Clustering & CAOM**: **Figure 13(a)**) and another example with one units on a channel (**Supplementary Material**: **Clustering & CAOM**: **Figure 13(b)**).

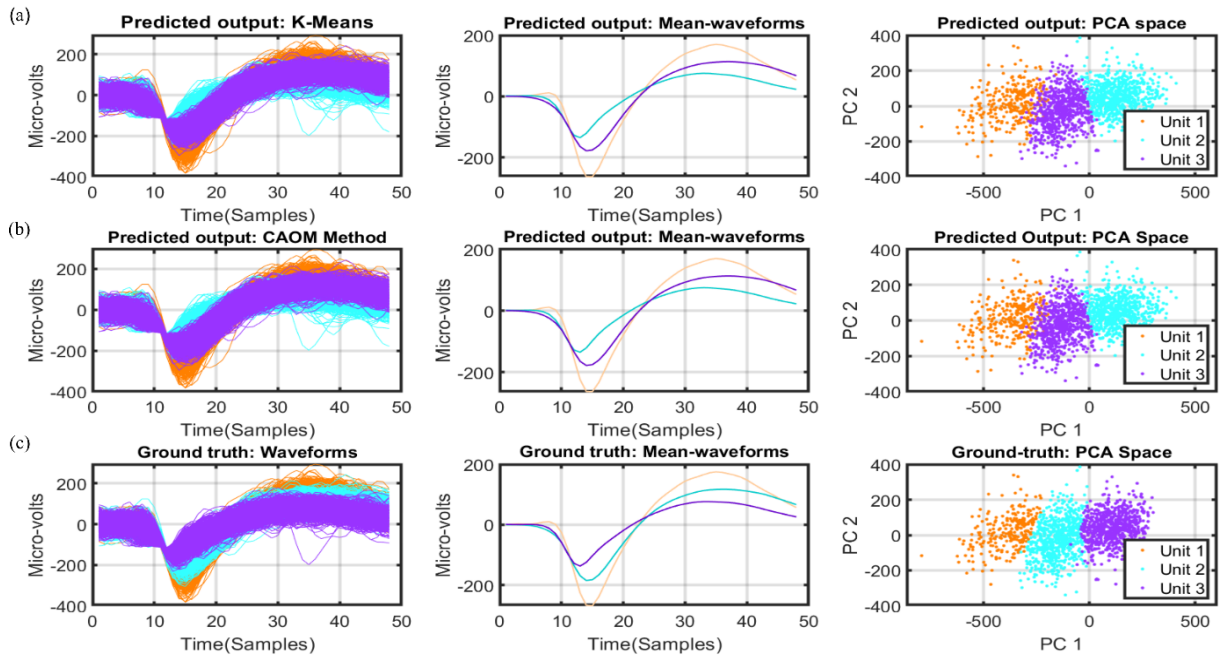

*Figure 7: Clustering algorithm in conjunction with CAOM (Visualization). (a) Output of K-Means clustering algorithm with pre-defined number of clusters is equals to 3. (b) Output of CAOM algorithm. Here, according to defined criteria of CAOM all three clusters were found to be distinct. (c) shows event with ground-truth labels. Predicted labels and ground truth looks quite similar.*

Similarly, **Figure 8** shows also provide visual insight of SpikeDeep-Classifier pipeline. **Figure 8** shows the output (projection in 2-D) of clustering algorithm, CAOM and the ground truths. **Figure 8** (a) shows an example of three neural unit on a channel, **Figure 8**(b) shows an example with two neural units on a channel and **Figure 8** (c) shows an example of a channel with one neural unit on it. In all the mentioned cases SpikeDeep-Classifier pipeline has not only been able to predict correct number of neural units on a channel but also predications are quite like ground truths.

**Table 4**, **Table 9**, **Figure 7**, **Figure 8** and **Figure 13** show that SpikeDeep-Classifier provides a reliable solution to spike sorting problem.

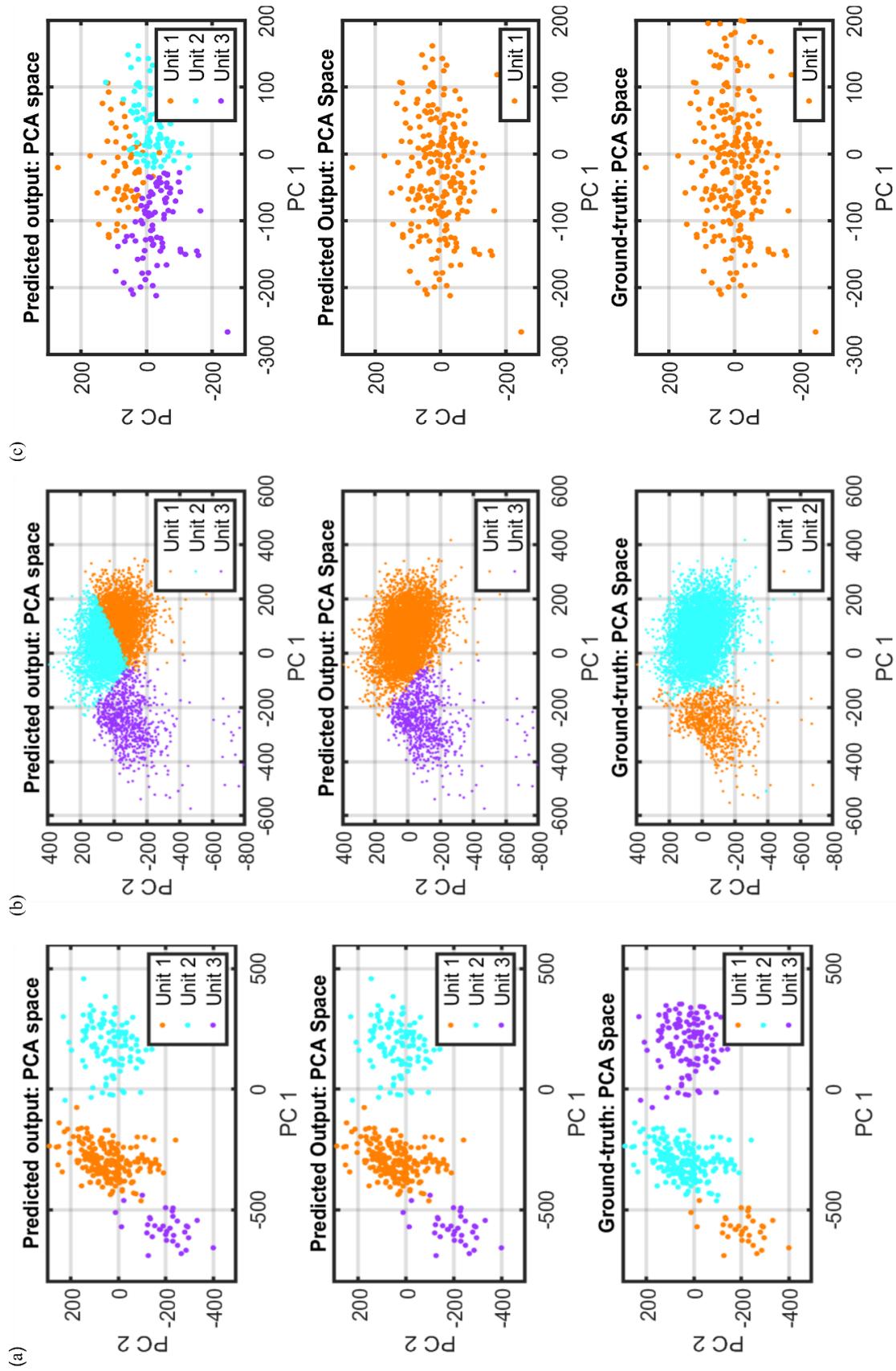

***Figure 8: Examples of Clustering and CAOM (Visualization).*** *(a) 2-D projection of events with predicted and gorund-truth labels of a channel with three distinct units. (b) ) 2-D projection of events with predicted and gorund-truth labels of a channel with two distinct units. (c) 2-D projection of events with predicted and gorund-truth labels of a channel with two distinct units. First row of each examples showsn the output of k-mean clustering algorithm, second row shows the output of CAOM and third row shows the ground truth label of the given events.*

# Clustering & CAOM: NHP Utah Array

We aimed for a universal solution to spike sorting problem. For that reason, we evaluated our trained models on data of multiple sessions of same species (humans), but multiple subjects implanted with different kinds of electrodes. Additionally, we evaluated same trained model for another species (NHPs), of multiple recording sessions of different subjects implanted with different kinds of electrodes and different recording hardware.

In this section, we will discuss evaluation and performance of Clustering & CAOM on the data of an NHP implanted with Utah arrays. We kept the parameters of clustering algorithm and CAOM fixed (number of clusters=3, similarity threshold = 5.5). In the recording session of NHP there is one channel with four neural units (see **Table 5**). In that case, our Clustering and CAOM has been able to predict 3 clusters with rand index= 0.78 and classification accuracy of 84.24%. However, number of clusters on this channel is debatable (see **Supplementary Material**: **Clustering & CAOM: Figure 14(f)** for visualization). Here, there are more wrong predications in term number of units on a channel. Therefore, we provided visual details of all the channels predicted with wrong predictions (number of clusters (see **Figure 9** and **Figure 14**)). **Figure 9** and **Figure 14** show that number of clusters on these channels are either debatable (See **Figure 9** (c, d, f) & **Figure 14** (b, d, e, f)) or wrongly labeled (**Figure 9** 8 (a, b, e) & **Figure 14** (a)). However, Rand index and classification accuracy remain consistent (see **Table 5**).

**Table 5** shows the result of SpikeDeep-Classifier of one Utah array implanted on PMd. For the same recording session another Utah array was implanted on M1. We reported the classification accuracy and Rand index along with number of units on a channel in **Table 10**.

*Table 5: Performance evaluation of clustering method & CAOM on the recording session of an NHP implanted with Utah array on PMd.*

| Number of Units | No. of Channels (True) | No. of Channels (Predicted) | Rand Index | Accuracy (%) |
|---|---|---|---|---|
| 4 | 1 | (0, 1) | 0.78 | 84.24 |
| 3 | 7 | (5, 3) | 0.86 ± 0.10 | 84.70 ± 11.79 |
| 2 | 25 | (22,8) | 0.88 ± 0.11 | 90.06 ± 10.20 |
| 1 | 18 | (12, 0) | 0.84± 0.18 | 89.02 ± 13.20 |

**Visualization:** We showed few correctly classified (number of clusters on a channel) examples in section (Results: Clustering & CAOM: Visualization). Here, we showed few wrongly classified examples for visual inspection (see **Figure 9**). Most of these examples are either debatable or wrongly labeled about the number units on a channel.

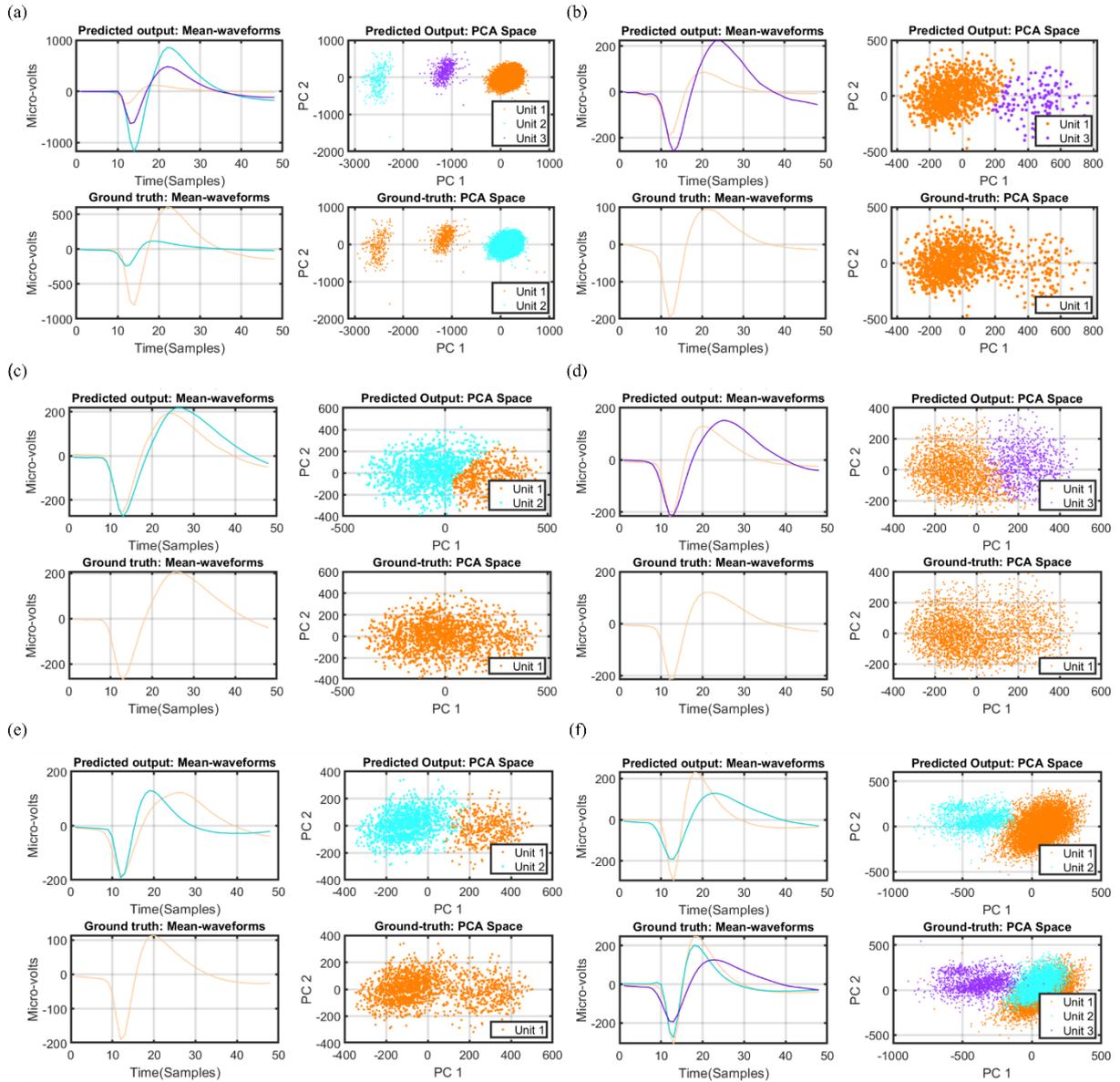

*Figure 9: Examples: Wrong classification in terms of number of clusters on a channel. For each example, first row shows mean waveforms of predicted clusters and 2-D projections of event with predicted labels., second row show the mean of each cluster and 2-D projection of events of assigned labels (ground-truth).*

## Clustering & CAOM: NHP Single Micro-electrode

We showed SpikeDeep-classifier provides a reliable solution to spike sorting problem. We evaluated on a data of human subjects with multiple recording sessions using different kinds of electrodes (see **Table 4** and table 9). Additionally, we evaluated it on a data of an NHP implanted with Utah on two different areas of brain (**Table 5** and Table 10). In addition, we evaluated it on a data of another NHP implanted with single micro electrode. Here, we did not have ground truth. Therfore, we showed performance of SpikeDeeep-classifier pipeline by presenting examples visually. We selected one example of each case. **Figure 10** (a) shows an example of a channel with one neural unit, **Figure 10** (b) shows an example of a channel with two neural units on a channel, and **Figure 10** (c) shows an example of a channel with three units.

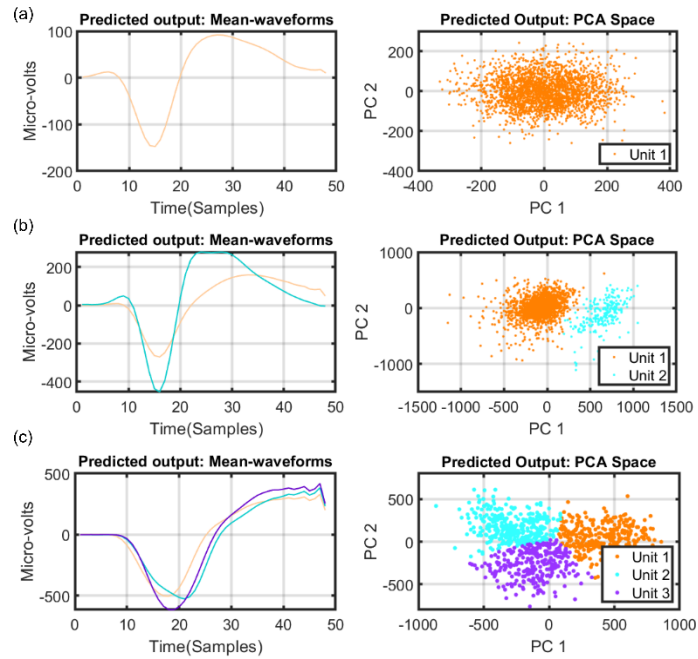

*Figure 10: **Visualization of clustering & CAOM of few channels (NHP implanted with single microelectrode recorded using Plexon).** Here, we did not have ground truth. However, results (visual inspection) show that SpikeDeep-classifier performs good quality of generalization. (a) Example with one neural unit. (b) Example with two-neural units. (c) Example with three neural units.*

# Supplementary Material

In this section, we provided additional information about to show generalization quality of SpikeDeep-Classifier pipeline. This section provides additional information (results) of each element of SpikeDeep-Classifier pipeline individually.

## SpikeDeeptector

In result section, we showed the cumulative performance of SpikeDeeptector on all the selected recording sessions. However, it is possible that performance of SpikeDeeptector is inconsistent during different recording. Therefore, we evaluated it on each selected session individually. Evaluation performance of SpikeDeeptector is shown in a table (see **Table 6**). SpikeDeeptector showed consistent performance during each recording sessions individually. **Table 6** also include the recording sessions of NHPs.

*Table 6: Supplementary material: Evaluation of SpikeDeeptector on the data recorded from all human patients and NHPs of all recording sessions, individually.*

| (Subject, Session) | (Spike channel/ Total Channel) | (Artifact channel/ Total Channel) | False Positives | False Negatives |
|---|---|---|---|---|
| (U1, S1) | 8/96 | 88/96 | 0 | 0 |
| (U1, S2) | 14/96 | 82/96 | 1 | 0 |
| (U1, S3) | 26/96 | 70/96 | 0 | 0 |
| (U2, S1) | 4/96 | 92/96 | 0 | 0 |
| (U2, S2) | 24/96 | 72/96 | 0 | 0 |
| (U2, S3) | 33/96 | 63/96 | 1 | 0 |
| (M1, S1) | 15/16 | 1/16 | 0 | 0 |
| (MM, PMD) | 51/51 | - | 0 | 0 |
| (MM, M1) | 44/45 | - | 0 | 1 |
| (X/B, PPC) | 4/4 | - | 0 | 0 |

## Non-neural Rejector (NNR)

In result section, we showed cumulative performance of NNR on all the selected. One valid question one could ask is, proof consistency of NNR. Therefore, in this section we show the consistency of NNR across multiple sessions of same subject and across different subjects. **Table 7** shows classification accuracy in confusion matrix of six recording session of two patients implanted with UTAH arrays, three recording sessions from each subject. Similarly, **Table 7** shows the classification accuracy of NNR across another human patient implanted with microwires. **Table 7** and **Table 8** shows the performance of NNR remains consistent across multiple recording sessions of different subjects, individually.

*Table 7:* Supplementary Material: Evaluation performance of NNR on the data recorded from both human patients of all six recording sessions, implanted with Utah arrays. Each confusion matrix reports overall accuracy and the classification accuracy of each class of each session.

**Subject: U2**

Session: 1

|  | Neural | Non-neural |  |
|---|---|---|---|
| Neural | 22450 / 49.9% | 2506 / 5.6% | 90.0% |
| Non-neural | 1617 / 3.6% | 18428 / 40.9% | 91.9% |
|  | 93.3% | 88.0% | **90.8%** |

Session: 2

|  | Neural | Non-neural |  |
|---|---|---|---|
| Neural | 75290 / 72.5% | 3389 / 3.2% | 95.7% |
| Non-neural | 8228 / 7.8% | 17872 / 17.6% | 68.5% |
|  | 90.1% | 84.1% | **88.9%** |

Session: 3

|  | Neural | Non-neural |  |
|---|---|---|---|
| Neural | 53458 / 78.7% | 1909 / 4.3% | 96.6% |
| Non-neural | 2420 / 3.6% | 10174 / 13.5% | 80.1% |
|  | 95.7% | 84.2% | **93.6%** |

**Subject: U1**

Session: 1

|  | Neural | Non-Neural |  |
|---|---|---|---|
| Neural | 5225 / 71.5% | 286 / 3.9% | 94.8% |
| Non-neural | 351 / 4.8% | 1444 / 19.8% | 80.5% |
|  | 93.7% | 83.5% | **91.3%** |

Session: 2

|  | Neural | Non-neural |  |
|---|---|---|---|
| Neural | 26745 / 87.5% | 423 / 1.4% | 98.4% |
| Non-neural | 994 / 3.3% | 2410 / 7.9% | 71.0% |
|  | 96.4% | 85.1% | **95.4%** |

Session: 3

|  | Neural | Non-neural |  |
|---|---|---|---|
| Neural | 22952 / 81.9% | 154 / 0.5% | 99.3% |
| Non-neural | 1873 / 6.7% | 3041 / 10.9% | 62.0% |
|  | 92.5% | 95.2% | **92.8%** |

Predicted Labels (y-axis) / True Labels (x-axis)

*Table 8:* Supplementary Material: Evaluation performance of NNR on the data recorded from a human patient of all one recording session, implanted with microwires. Confusion matrix reports overall accuracy and the classification accuracy of each class.

**Subject: M2**

|  | Neural | Non-neural |  |
|---|---|---|---|
| Neural | 178250 / 85.3% | 2362 / 1.1% | 97.9% |
| Non-neural | 11558 / 5.5% | 16727 / 8.0% | 59.1% |
|  | 93.3% | 87.6% | **93.3%** |

Predicted Labels / True Labels

## Significance: Non-neural Rejector

Most spike sorting algorithms performs clustering on two-dimensional projected space. However, it is possible that useful information lost, during dimensionality reduction. As a result, discrimination between non-neural activity and neural activity is affected (see **Figure 11**(a) & **Figure 12**(a)). Non-neural cluster and spike cluster completely overlap. Therefore, it is hard for any clustering algorithm to discriminate multiple clusters here. During manual spike sorting, in **Figure 11** (a) even a human made mistake and merged the non-neural activity cluster and spike cluster. However, NNR enable us to isolate these two clusters (see **Figure 11**(a)), which later provide an ease to clustering algorithm and CAOM as shown in **Figure 11** (b). Therefore, we considered NNR as a preprocessing step for clustering method. **Figure 12** show another example, which shows the significance of NNR. We believe including NNR as a preprocessing step for spike sorting, makes it an ordinary clustering problem.

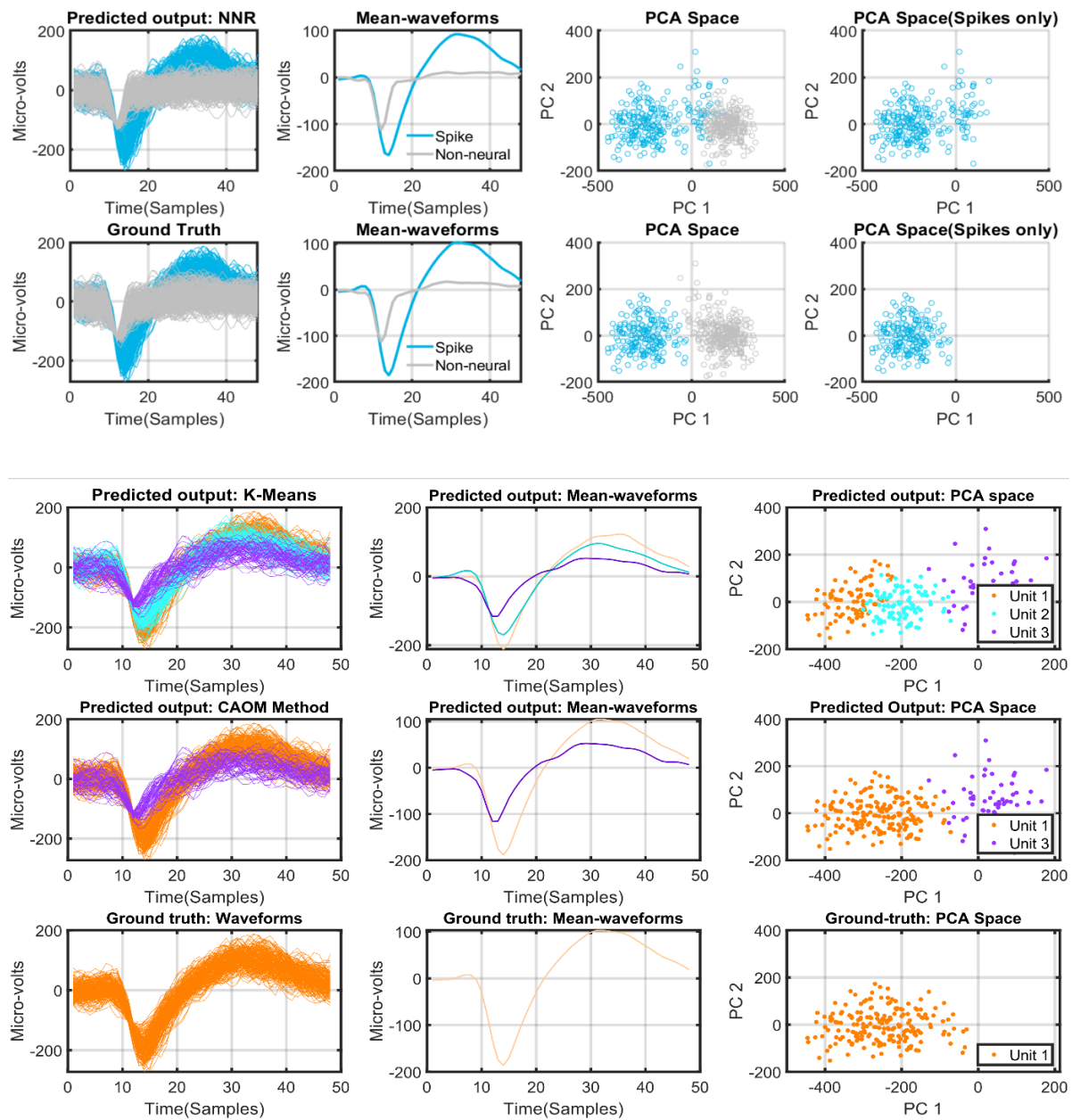

*Figure 11: Supplementary Material: Significance of NNR (Visualization)* (a) Response of NNR along with ground truth labels (human). NNR has clearly outperformed human in this case. Here, non-neural cluster and spike cluster was completely overlapped. Therefore, even human misclassify some of the events and can missed a neural unit. (b) Shows the result of clustering and CAOM. SpikeDeep-Classifier pipeline outperforms human spike sorter because NNR isolates overlapped

*clusters. Mean waveforms of both clusters (see Result CAOM) clearly show two distinct neural units. However, human curator missed one cluster (see ground-truth).*

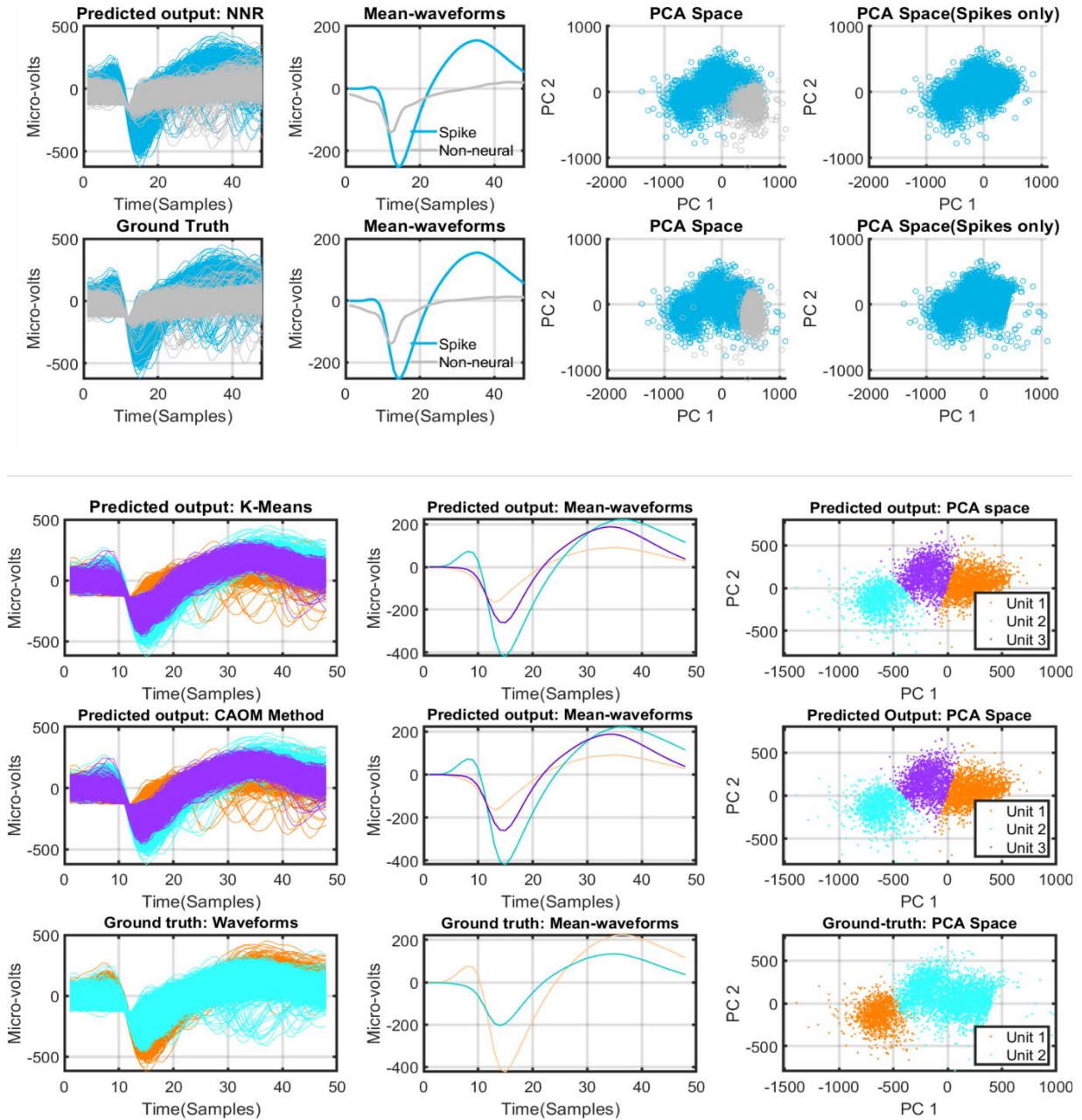

*Figure 12: Supplementary Material: Significance of NNR (Visualization) (a) Response of NNR along with ground truth labels (human). Here, non-neural cluster and spike cluster was completely overlapped. Performance of NNR is comparable to human. (b) Shows the result of clustering and CAOM. SpikeDeep-Classifier pipeline outperforms human spike sorter because NNR isolates overlapped clusters. Mean waveforms of all three clusters (see Result CAOM) clearly show three distinct neural units. However, human curator missed one cluster (see ground-truth).*

## Clustering & CAOM

We showed SpikeDeeptector and NNR shows good quality of generalization across multiple recording sessions of different subjects implanted with different types of electrodes. In this section, we showed consistency of clustering & CAOM across multiple sessions of humans and NHPs. We used the fixed value of parameters, number of clusters=3, similarity threshold = 5.5.

**Utah Array Human Subjects:** We used same recording sessions, previously used for the evaluation consistency of SpikeDeeptector and NNR. Table 9 shows that performance of Clustering algorithm and CAOM remain consistent, during each individual session of human patients implanted with Utah array. The selected sessions have different number of channels with varying number of neural unit on a channels e.g., (U2, S2) & (U2, S3) have more channels with three neural units on channels, and (U1, S2) & (U1, S3) have more channels with one unit on a channel. However, performance of Clustering and CAOM remains consistent during all kinds of recording sessions (see **Table 9**).

*Table 9: Supplementary Material:* *Performance evaluation of clustering method & CAOM on the data of six recording sessions of human patients implanted with Utah arrays, individually.*

| No. of units | (Subject, Session) | No. of Channels (True) | No. of Channels (Predicted) (Correct, wrong) | Rand Index (%) |
|---|---|---|---|---|
| 3 | (U1, S1) | 1 | (1, 0) | 0.843 |
|   | (U1, S2) | 0 | (0, 0) | - |
|   | (U1, S3) | 0 | (0, 0) | - |
|   | (U2, S1) | 0 | (0, 1) | - |
|   | (U2, S2) | 9 | (9, 0) | 0.8 ± 0.07 |
|   | (U2, S3) | 8 | (8, 0) | 0.89 ± 0.06 |
| 2 | (U1, S1) | 4 | (3, 1) | 0.83 ± 0.17 |
|   | (U1, S2) | 2 | (2,1) | 0.89 ±0.04 |
|   | (U1, S3) | 7 | (7, 1) | 0.88 ± 0.07 |
|   | (U2, S1) | 4 | (3, 0) | 0.85 ± 0.08 |
|   | (U2, S2) | 11 | (11, 2) | 0.86 ± 0.06 |
|   | (U2, S3) | 19 | (18, 0) | 0.82 ± 0.11 |
| 1 | (U1, S1) | 3 | (2, 1) | 0.88 ± 0.90 |
|   | (U1, S2) | 11 | (10, 0) | 0.80 ± 0.12 |
|   | (U1, S3) | 19 | (18, 0) | 0.81 ± 0.16 |
|   | (U2, S1) | 0 | (0, 0) | - |
|   | (U2, S2) | 4 | (2, 0) | 0.78 ± 0.05 |
|   | (U2, S3) | 5 | (5, 1) | 0.85 ± 0.05 |

**Visualization:** We provided few additional examples for visualization. These examples show quality of spike sorting provided by SpikeDeep-classifier algorithm in comparison with ground truth labels. **Figure 13** shows an example of a channel with one neural unit and another example with two neural unit.

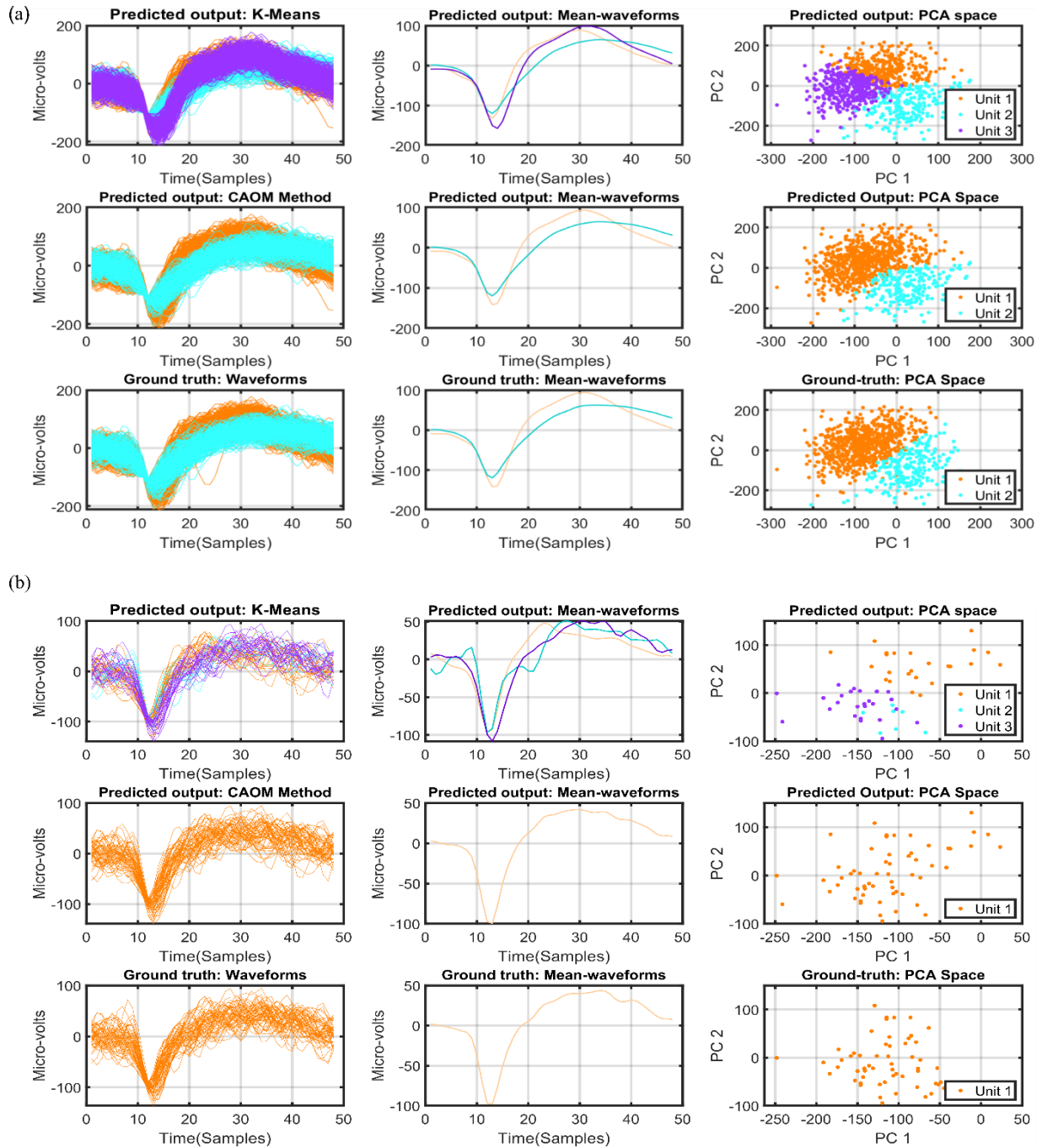

*Figure 13: Supplementary Material: Clustering algorithm in conjunction with CAOM (Visualization). (a) An example with two neural units on a channel. First row: Output of K-Means clustering algorithm with pre-defined number of clusters is equals to 3. Second row: Output of CAOM algorithm. Here, according to defined criteria of CAOM two are clusters merged and one remaining cluster is considered as a distinct cluster. Third row: shows event with ground-truth labels. Predicted labels and ground truth looks quite similar. (b) An example with one neural unit on a channel First row: Output of K-Means clustering algorithm with pre-defined number of clusters is equals to 3. Second row: Output of CAOM algorithm. Here, according to defined criteria of CAOM all three clusters are merged. Third row: shows event with ground-truth labels. Predicted labels and ground truth looks quite similar.*

**Utah Array NHPs:** In addition to human patients, we also evaluated clustering & CAOM method on the data of an NHP to check the generalization quality and consistency. **Table 10** shows that performance of Clustering algorithm in conjunction with CAOM remains consistent. Similar performance of clustering and CAOM is reported for the same recording session of a same subject but different area of brain (see **Table 5**).

*Table 10: **Supplementary Material:** Performance evaluation of clustering method & CAOM on the recording session of an NHP implanted with Utah array on M1.*

| Number of Units | No. of Channels (True) | No. of Channels (Predicted) | Rand Index | Accuracy (%) |
| --- | --- | --- | --- | --- |
| 4 | 1 | (0, 0) | 0.74 | 65.02 |
| 3 | 2 | (1, 1) | 0.80 ± 0.18 | 79.45 ± 22.29 |
| 2 | 14 | (14, 7) | 0.96 ± 0.03 | 97.5± 2.4 |
| 1 | 28 | (22, 0) | 0.90± 0.13 | 94.3± 8.4 |

**Visualization:** We showed few correctly classified (number of clusters on a channel) examples in section (**Supplementary material: Clustering & CAOM**: **Utah Array Human Subjects: Visualization**). Here, we showed remaining wrongly classified examples for visual inspection of recording session (see **Figure 14**) of recording session (MM, PMd). Most of these examples are either debatable or wrongly labeled about the number units on a channel.

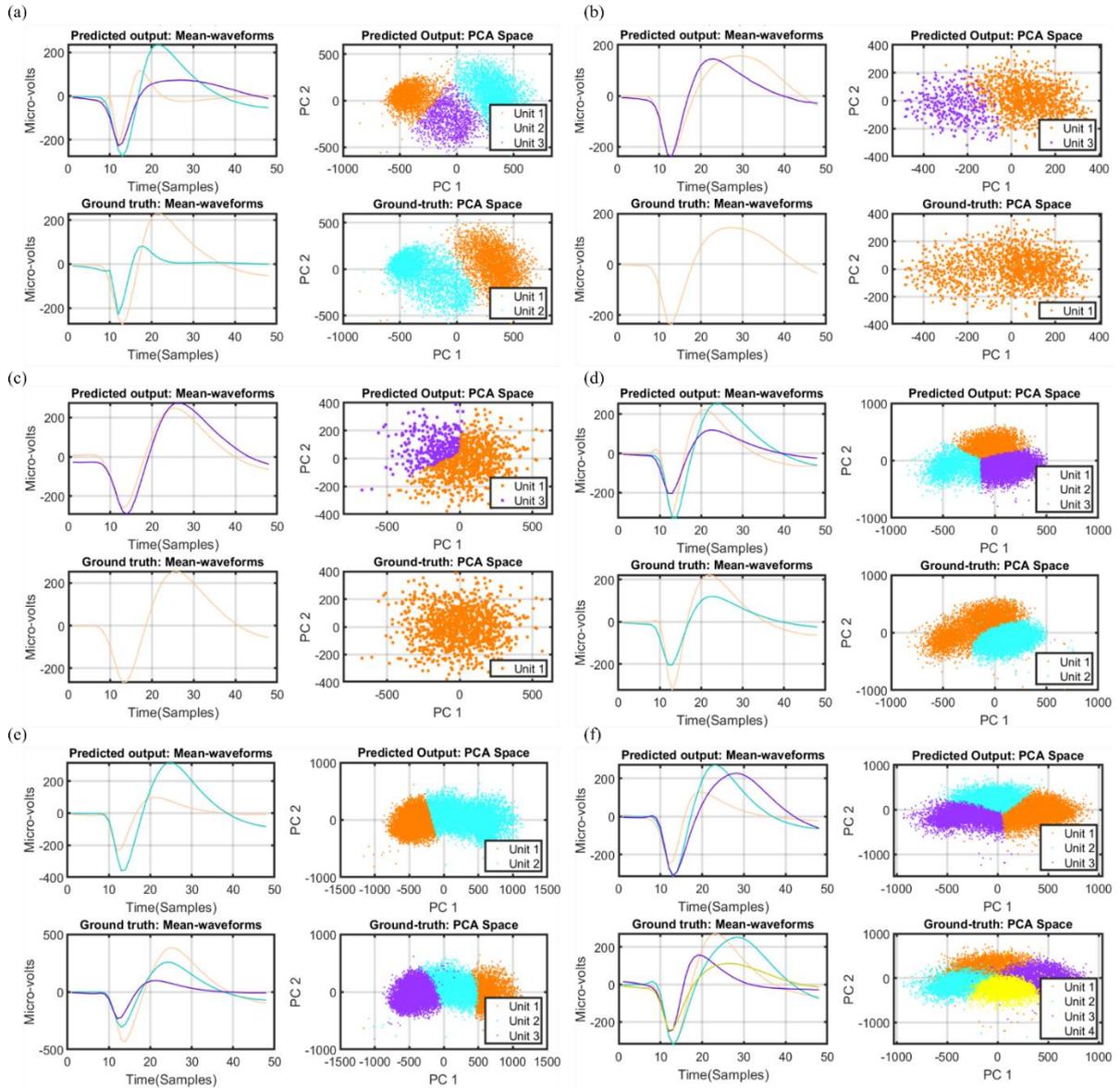

*Figure 14: Supplementary Material: Remaining Examples: Wrong classification in terms of number of clusters on a channel.* For each example, first row shows mean waveforms of predicted clusters and 2-D projections of event with predicted labels., second row show the mean of each cluster and 2-D projection of events of assigned labels (ground-truth).